\PassOptionsToPackage{linktocpage}{hyperref}
\documentclass[twocolumn,twocolappendix,tighten]{aastex6}
\usepackage{amsmath}

\begin{document}


\title{Association of $^3$H\lowercase{e}-Rich Solar Energetic Particles with Large-Scale Coronal Waves}



\author{Radoslav Bu\v{c}\'ik\altaffilmark{1,2} }
\affil{Institut f\"{u}r Astrophysik, Georg-August-Universit\"{a}t G\"{o}ttingen, D-37077, G\"{o}ttingen, Germany}


\author{Davina E. Innes\altaffilmark{3}}
\affil{Max-Planck-Institut f\"{u}r Sonnensystemforschung, D-37077, G\"{o}ttingen, Germany}


\author{Glenn M. Mason}
\affil{Applied Physics Laboratory, Johns Hopkins University, Laurel, MD 20723, USA}


\and


\author{Mark E. Wiedenbeck}
\affil{Jet Propulsion Laboratory, California Institute of Technology, Pasadena, CA 91109, USA}



\altaffiltext{1}{Max-Planck-Institut f\"{u}r Sonnensystemforschung, D-37077, G\"{o}ttingen, Germany}
\altaffiltext{2}{bucik@mps.mpg.de}
\altaffiltext{3}{Max Planck/Princeton Center for Plasma Physics, Princeton, NJ 08540, USA}

\begin{abstract}

Small $^3$He-rich solar energetic particle (SEP) events have been commonly associated with extreme-ultraviolet (EUV) jets and narrow coronal mass ejections (CMEs) which are believed to be the signatures of magnetic reconnection involving\deleted{open}field \added{lines open to interplanetary space}. The elemental and isotopic fractionation in these events are thought to be caused by processes confined to the flare sites. In this study we identify 32 $^3$He-rich SEP events observed by the {\sl Advanced Composition Explorer} near the Earth during the solar minimum period 2007--2010 and examine their solar sources with the high resolution {\sl Solar Terrestrial Relations Observatory} (STEREO) EUV images. Leading the Earth, STEREO-A provided for the first time a direct view on $^3$He-rich flares, which are generally located on the Sun's western hemisphere. Surprisingly, we find that about half of the $^3$He-rich SEP events in this survey are associated with large-scale EUV coronal waves. An examination of the wave front propagation, the source-flare distribution and the coronal magnetic field \replaced{extrapolations}{connections} suggests that the EUV waves may affect the injection of $^3$He-rich SEPs into interplanetary space.

\end{abstract}

\keywords{acceleration of particles --- shock waves --- Sun: coronal mass ejections (CMEs) --- Sun: flares --- Sun: particle
emission --- waves}



\section{Introduction} \label{sec:intro}

Solar energetic particles (SEPs) can be accelerated either at coronal mass ejection (CME) driven shocks in gradual large SEP (LSEP) events or by processes related to magnetic reconnection at the flare sites in impulsive or $^3$He-rich SEP events \citep[see review by][]{rea13}. Anomalous elemental composition in impulsive SEP events, extremely different from coronal abundances, has not been so far sufficiently explained \citep[see][for a review]{koc84,mas07,rea15}. In addition to a preferential acceleration by resonant interactions with MHD turbulence \citep[e.g.,][]{mil98,liu06}, other non-wave mechanisms have been proposed \citep[e.g.,][]{dra12}. In contrast to LSEP events, $^3$He-rich SEP events have lower ion intensities and energies, shorter durations and are associated with minor X-ray flares. Though they are numerous these features make them a less explored phenomenon.   

Solar sources of $^3$He-rich SEPs have been often associated with coronal X-ray and EUV jets \citep[e.g.,][]{nit06,nit08,wan06,buc14,che15,inn16} which sometimes show a high altitude extension in jet-like CMEs \citep{kah01,wan06,wan12}. Jets have been considered to be a signature of magnetic reconnection involving open field \citep[e.g.,][]{shi92}. Predominantly narrow CMEs in these events have been also reported by \citet{nit06}. Recently, new observations from the {\sl Solar Dynamics Observatory} (SDO) and  {\sl Solar Terrestrial Relations Observatory} (STEREO) have revealed a few $^3$He-rich SEP events associated with large-scale coronal EUV waves \citep{nit15,buc15}. This finding appears to be surprising in view of previous associations with narrow flare signatures.

Four of the reported $^3$He-rich SEP events with a coronal wave were accompanied by slow ($\lesssim$300\,km\,s$^{-1}$) CMEs \citep{nit15} but two others were without a CME \citep{buc15}. Coronal waves have been so far reported in LSEP events in association with fast and wide CMEs \citep[e.g.,][]{tor99,lar14,mit14}. However, there is still some controversy on the nature of coronal waves \citep[see][and references therein]{pat12,liu14,war15}. Presently, it is believed they are true magnetosonic waves and not just coronal magnetic field reconfigurations induced by a CME. Indeed, a recent report on EUV waves associated with X-ray flares alone \citep{nit13} appears to be consistent with the true wave concept. 

The aim of this paper is to examine how common an association of $^3$He-rich SEPs with large-scale coronal waves is without setting any restriction criteria on the selection of the events. For this purpose we collect 32 $^3$He-rich SEP events within the solar activity minimum period 2007--2010 to avoid source confusion in the more active periods. We identify solar sources of $^3$He-rich SEPs using new viewing perspectives from angularly separated spacecraft. These observations are presented in Section \ref{sec:rad}. In Section \ref{sec:con} we discuss possible implications of associated coronal waves on energetic ion characteristics in these events.

\section{Results} \label{sec:rad}

\subsection{$^3$He-rich SEP events} \label{subsec:ev}

Table 1 lists 32 $^3$He-rich SEP events observed near the Earth at the L1 point with the {\sl Advanced Composition Explorer} (ACE) between January 2007 and December 2010. We have selected all detectable events by examining He mass spectrograms of low-energy (0.2--1\,MeV\,nucleon$^{-1}$) ions from the Ultra Low Energy Isotope Spectrometer \citep[ULEIS;][]{mas98} and of high-energy (4.5--16.3\,MeV\,nucleon$^{-1}$) ions from the Solar Isotope Spectrometer \citep[SIS;][]{sto98} including the events near the detection threshold reported in the previous studies \citep{mas09,wie13}. The ULEIS is a high-resolution time-of-flight mass spectrometer which measures ions from He to Ni in the energy range from 20\,keV\,nucleon$^{-1}$ to several MeV\,nucleon$^{-1}$.  The SIS is a $dE/dx$ versus residual energy telescope measuring He to Zn nuclei from $\sim$10 to $\sim$100\,MeV\,nucleon$^{-1}$ range. Only 1/4 of all examined events showed a clear signature of $^3$He at the highest examined energy range of 7.6--16.3\,MeV\,nucleon$^{-1}$. The high-energy $^3$He-rich SEPs are relatively rare events \citep{les03} and it is not clear if they require an additional acceleration process \citep{wie10a}.

\begin{deluxetable*}{clccccccc}
\tablecaption{L1 $^3$He-rich SEP events Jan 2007--Dec 2010: Energetic ion properties \label{tab:tab1}}
\tablehead{
\colhead{Number} & \colhead{Start date} & \colhead{Days\tablenotemark{a}} & \colhead{$^3$He fluence\tablenotemark{b}} & \colhead{$^3$He/$^4$He\tablenotemark{b}} & \colhead{Fe/O\tablenotemark{b}}  & \colhead{\added{Dispersion}} & \colhead{IMF} & \colhead{References}\\
\colhead{} & \colhead{} & \colhead{} & \colhead{($\times$10$^{3}$)} & \colhead{} & \colhead{} & \colhead{} & \colhead{polarity\tablenotemark{c}} & \colhead{}
}
\decimals
\startdata
1 & 2007 Jan 14 & \phn14.25 -- \phn14.54 & \phn1.22$\pm$0.29 & 0.14\phn$\pm$0.04\phn & 0.70$\pm$0.36 & no & ($-$) &\\
2 & 2007 Jan 24 & \phn24.50 -- \phn27.50 & \phn7.51$\pm$0.73 & 0.12\phn$\pm$0.01\phn & 1.04$\pm$0.11 & yes & (+) & 1\\
3 & 2007 May 23 & 143.31 -- 143.92 & \phn2.90$\pm$1.40 & 0.011$\pm$0.006 & 2.30$\pm$1.00 & no & (+) & 2\\
4 & 2008 Feb 4 & \phn35.00 -- \phn37.50 & \phn7.20$\pm$2.30 & 0.16\phn$\pm$0.05\phn & $\sim$1 & no & ($-$) & 2\\
5 & 2008 Jun 16 & 168.90 -- 170.00 & \phn1.15$\pm$0.32 & 0.12\phn$\pm$0.03\phn & \nodata & ? & ($-$) & 2\\
6 & 2008 Nov 4 & 309.50 -- 312.71 & \phn1.44$\pm$0.37 & 0.05\phn$\pm$0.01\phn & 1.45$\pm$0.44 & yes & ($-$) & 1, 2, 3\\
7 & 2009 Apr 29 & 119.90 -- 121.20 & \phn0.96$\pm$0.32 & 0.90\phn$\pm$0.41\phn & \nodata & ? & (+) & 3\\
8 & 2009 May 1 & 121.70 -- 122.50 & \phn0.64$\pm$0.26 & 1.50\phn$\pm$0.97\phn & \nodata & ? & ($-$) & \\
9 & 2009 Jul 5 & 186.70 -- 189.50 & \phn0.33$\pm$0.19 & 0.33\phn$\pm$0.22\phn & \nodata & ? & ($-$) & 3\\
10 & 2009 Oct 6 & 279.88 -- 281.75 & \phn1.64$\pm$0.44 & 1.40\phn$\pm$0.58\phn & \nodata & ? & ($-$) & \\
11 & 2009 Nov 3   & 307.25 -- 308.75   &    \nodata     & \nodata &             \nodata & ? & ($-$) & 3\\
12 & 2009 Dec 22 & 356.88 -- 359.50   &   \phn1.55$\pm$0.43 & 0.23\phn$\pm$0.07\phn & 1.16$\pm$0.62 & no & (+) & \\
13 & 2010 Jan 16 &    \phn16.29 -- \phn18.00  &   \phn4.29$\pm$0.72 & 1.64\phn$\pm$0.44\phn &  \nodata & yes & ($-$) & \\
14 & 2010 Jan 27 &    \phn27.13 -- \phn28.63   &   \phn5.72$\pm$0.83 & 0.13\phn$\pm$0.02\phn & 0.58$\pm$0.10 & yes & (+) & 4\\
15 & 2010 Jan 31 &    \phn31.00 -- \phn32.00   &  \phn0.95$\pm$0.34 & 4.00\phn$\pm$3.16\phn &   \nodata & ? & ($-$) & \\
16 & 2010 Feb 8   &    \phn39.66 -- \phn42.50   & 21.5\phn$\pm$1.6\phn  & 0.21\phn$\pm$0.02\phn & 0.89$\pm$0.16 & no & ($-$) & 3\\
17 & 2010 Feb 12 & \phn43.75 -- \phn45.33 & \phn7.28$\pm$0.94 & 0.018$\pm$0.002 & 0.35$\pm$0.04 & no & ($-$) & 3\\
18 & 2010 Feb 19 & \phn50.08 -- \phn50.33 & \phn7.87$\pm$0.98 & 0.52\phn$\pm$0.08\phn & 1.16$\pm$0.24 & yes & (+) & 3\\
19 & 2010 Mar 4 & \phn63.79 -- \phn64.50 & \phn0.95$\pm$0.34 & 0.033$\pm$0.012 & 0.73$\pm$0.24 & yes & ($-$) & \\
20 & 2010 Mar 19 & \phn78.58 -- \phn79.58 & \phn7.03$\pm$0.92 & 0.37\phn$\pm$0.06\phn & 1.32$\pm$0.38 & yes & (+) & \\
21 & 2010 Jun 12 & 163.75 -- 166.13 & 22.9\phn$\pm$1.7\phn & 0.023$\pm$0.002 & 0.94$\pm$0.08 & no & ($-$) & 3\\
22 & 2010 Sep 1 & 244.25 -- 244.83 & \phn1.31$\pm$0.40 & 0.021$\pm$0.006 & 0.75$\pm$0.21 & no & ($-$) & 3\\
23 & 2010 Sep 2 & 245.75 -- 247.08 & \phn7.76$\pm$0.96 & 3.10\phn$\pm$0.78\phn & 0.87$\pm$0.67 & no & ($-$) & \\
24 & 2010 Sep 4 & 247.63 -- 248.88 & \phn9.95$\pm$1.09 & 0.086$\pm$0.010 & 0.24$\pm$0.04 & yes & ($-$) & \\
25 & 2010 Sep 17 & 260.50 -- 261.75 & \phn0.48$\pm$0.24 & 0.034$\pm$0.017 & \nodata & no & ($-$) & \\
26 & 2010 Oct 17 & 290.83 -- 292.75 & 21.0\phn$\pm$1.6\phn & 0.46\phn$\pm$0.04\phn & 1.27$\pm$0.23 & yes & (+) & 3, 5, 6\\
27 & 2010 Oct 19 & 292.33 -- 293.00 & 14.4\phn$\pm$1.3\phn & 1.53\phn$\pm$0.22\phn & 1.93$\pm$0.43 & yes & ($-$) & 3, 5\\
28 & 2010 Nov 2 & 306.75 -- 308.75 & 23.5\phn$\pm$1.7\phn & 0.94\phn$\pm$0.09\phn & 0.91$\pm$0.32 & yes & (+) & 3, 5\\
29 & 2010 Nov 14 & 318.38 -- 321.46 & 26.3\phn$\pm$1.8\phn & 0.20\phn$\pm$0.01\phn & 1.20$\pm$0.18 & no & ($-$) & 3, 5, 6, 7\\
30 & 2010 Nov 17 & 321.67 -- 322.42 & 32.8\phn$\pm$2.0\phn & 2.75\phn$\pm$0.33\phn & 1.25$\pm$0.34 & yes & ($-$) & 5, 6, 7\\
31 & 2010 Nov 29 & 333.50 -- 334.25 & \phn3.81$\pm$0.67 & 0.91\phn$\pm$0.22\phn & \nodata & ? & (+) & \\
32 & 2010 Dec 10 & 344.71 -- 346.25 & \phn1.55$\pm$0.43 & 0.59\phn$\pm$0.21\phn & \nodata & ? & ($-$) & \\
\enddata
\tablenotetext{a}{at 200--400\,keV\,nuc.$^{-1}$ but event 3 from Ref. (2); event 11 at 4.5--7.6\,MeV\,nuc.$^{-1}$; events 17, 27 at 7.6--16.3\,MeV\,nuc.$^{-1}$; 19, 21, 22, 24, 31 at 0.4--1\,MeV\,nuc.$^{-1}$}
\tablenotetext{b}{at 320--450\,keV\,nuc.$^{-1}$; fluence unit -- particles (cm$^2$\,sr\,MeV/nuc.)$^{-1}$; values in events 3, 4 from Ref. (2); event 11 not seen at the ULEIS energies - an upper limit for $^3$He/$^4$He at the lowest SIS energy bin (5.2\,MeV\,nuc.$^{-1}$) is 0.35$+$0.16}
\tablenotetext{c}{($-$) negative and (+) positive polarity with IMF oriented toward and away from the Sun respectively}
\tablecomments{\added{Events 3, 6--9, 11, 12, 14--22, 26, 28, 31 and 32 are associated with EUV waves (see Table \ref{tab:tab2}).}}
\tablerefs{(1) \citet{wie10}; (2) \citet{mas09}; (3) \citet{wie13}; (4) \citet{buc15}; (5) \citet{nit15}; (6) \citet{buc13}; (7) \citet{che15}.}
\end{deluxetable*}

Column 1 in Table \ref{tab:tab1} indicates the event number, column 2 the particle event start date. Column 3 shows the approximate $^3$He-rich interval at 200--400\,keV\,nucleon$^{-1}$ (in decimal days) where the ion travel time from the Sun is $\sim$7 hr without scattering. Sometimes we have used higher energy ranges to define the $^3$He period (see Table \ref{tab:tab1}) where the event was clearer. Columns 4, 5, 6 provide $^3$He fluence, $^3$He/$^4$He and Fe/O abundance ratios, respectively, at 320--452\,keV\,nucleon$^{-1}$ integrated over the period shown in column 3. \added{Column 7 indicates whether or not the event showed velocity dispersion at the onset.} Column 8 gives the interplanetary magnetic field (IMF) polarity during the event, determined from the observed IMF \citep{smi98} vector orientation relative to the nominal Parker magnetic field. The Parker angle was calculated using the observed solar wind speed \citep{mcc98}. The last column indicates references to previous studies of these events. Previously 19 events from this survey were reported but the solar source flare was identified only for 9 events. In 2007 and 2008 we found only 3 events per year.  An increase of sunspot numbers led to the 6 events in 2009. In 2010 we identified 20 events.  

\begin{figure*}
\figurenum{1}
\plotone{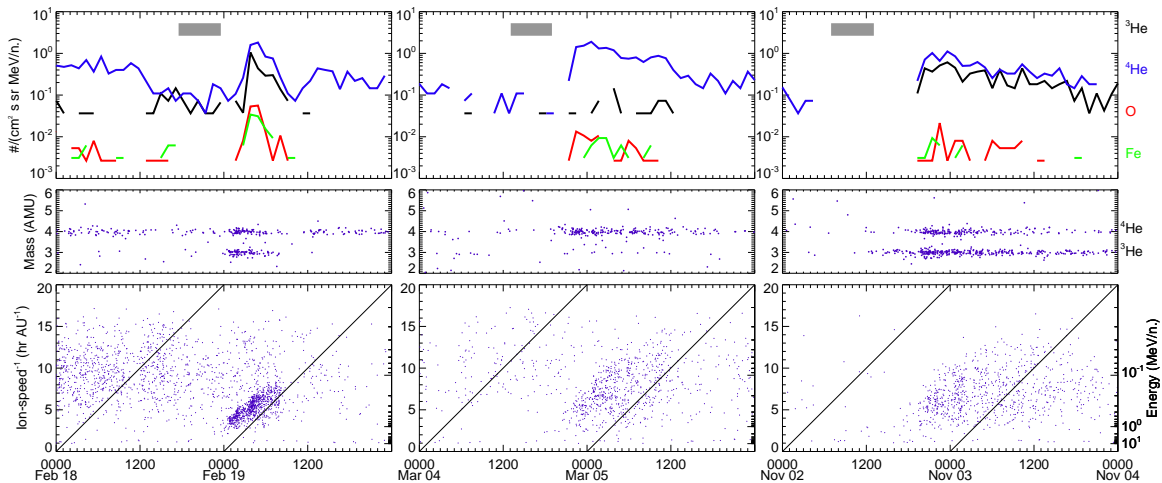}
\caption{Top: 1 hr ACE/ULEIS 230--320\,keV\,nucleon$^{-1}$ $^4$He, $^3$He, O, Fe intensities. The horizontal shaded bars indicate the 6 hr periods where solar electron events with type III radio bursts are shown in Figure \ref{fig:f2}. Middle: ULEIS 0.4--10\,MeV\,nucleon$^{-1}$ He mass spectrograms of individual pulse-height analysed ions. Bottom: ULEIS spectrograms of 1/(ion-speed) versus arrival times of 10--70\,amu ions. Sloped lines indicate arrival times for particles traveling along a field line of 1.2\,AU without scattering. \label{fig:f1}}
\end{figure*}

Four of the $^3$He-rich SEP events in this survey (numbers 12, 17, 21, 22) were accompanied by $>$25 MeV \citep[SOHO/ERNE;][]{tor95} lower-intensity solar proton events \citep{ric14}. These events probably include a mix of both $^3$He-rich and LSEP components. Such events with both flare and gradual components have been previously reported \citep[e.g.,][]{coh99,mas99,can03}. Two of these events (12, 21) have type II radio bursts in the Solar Radio Bursts Report (\href{ftp://ftp.ngdc.noaa.gov/STP/space-weather/solar-data/solar-features/solar-radio/radio-bursts}{ftp://ftp.ngdc.noaa.gov/STP/space-weather/solar-data} \href{ftp://ftp.ngdc.noaa.gov/STP/space-weather/solar-data/solar-features/solar-radio/radio-bursts}{/solar-features/solar-radio/radio-bursts}). The source for event 22 was limb occulted. The 386\,keV\,nucleon$^{-1}$ $^3$He/$^4$He ratio attains the lowest values in these events, $\sim$2\% in three events (17, 21, 22) but in event 12 the ratio reached 23\%. The events 17 and 22 are more enhanced at high energies than low. The 10.5\,MeV\,nucleon$^{-1}$ $^3$He/$^4$He ratio is $0.10\substack{+0.04 \\ -0.02}$ and $0.11\substack{+0.04 \\ -0.02}$ for events 17 and 22, respectively. The average 386\,keV\,nucleon$^{-1}$ Fe/O ratio in these four events is $\sim$80\%, typical of $^3$He-rich SEPs. All these events are associated with a coronal EUV wave (see Section \ref{subsec:ss}). Note that another event (number 16) occurred during the decay phase of a gradual SEP event. Half of the four $^3$He-rich SEP events ($^3$He/$^4$He $\sim$8\% and 36\% in 0.5--2.0\,MeV\,nucleon$^{-1}$) with coronal waves reported by \citet{nit15} were also accompanied by weak high-energy solar proton events. 

\deleted{The following $^3$He-rich SEP events in this survey were simultaneously observed on angularly separated STEREO-A (STA) or -B (STB) spacecraft \citep{wie10,wie13}: event 6 on STB ($-41^\circ$), 7 on STA ($+48^\circ$), event 9 with significantly higher intensity on STB ($-48^\circ$), 11- stronger on STA ($+63^\circ$), events 16 and 17 on STA ($+65^\circ$) and STB ($-71^\circ$), 21 on STB ($-70^\circ$), 22 on STA ($+81^\circ$), 28 on STA ($+84^\circ$). The values in parentheses indicate the spacecraft separation angle from the Earth. STA is leading and STB trailing the Earth on their heliocentric orbits at $\sim$1\,AU. The STEREO $^3$He-rich events were identified using the LET \citep{mew08} instrument in the range 2.3--3.3\,MeV\,nucleon$^{-1}$. The authors excluded chance coincidences between detections of $^3$He-rich SEPs from widely spaced active regions (ARs).}

\begin{figure}
\figurenum{2}
\plotone{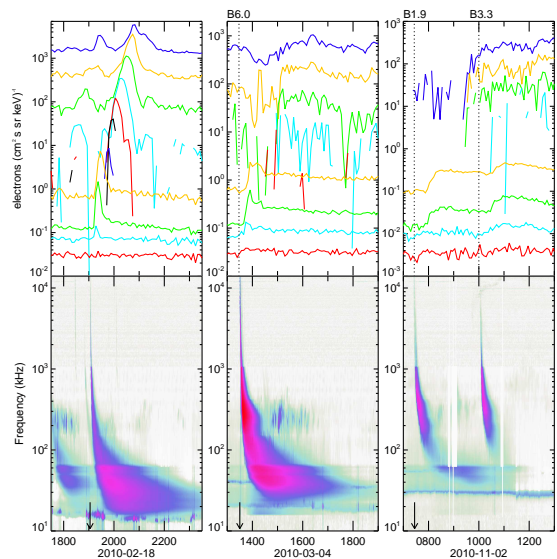}
\caption{Top: 5-minute electron intensities centered at 2, 2.8, 4.2, 6.1, 8.9, 13, 19, 45, 74, 134, 235\,keV (from the upper curve). In the 2010 November 2 event the last 4 energies are at 40, 66, 108, 182\,keV. Data at 2--19\,keV are from EESA-H Wind/3DP telescope and at 45--238\,keV from DE30 ACE/EPAM telescope. Data at 40--182\,keV are from SST Wind/3DP telescope and are not available for the 2010 February and March events. The dotted vertical lines mark the start times of the GOES X-ray flares. Bottom: The Wind/WAVES radio spectrograms. The arrows mark the start times of the type III bursts associated with the $^3$He-rich SEP events in Figure \ref{fig:f1}. \label{fig:f2}}
\end{figure}

Figure \ref{fig:f1} shows three examples of $^3$He-rich SEP events associated with a coronal EUV wave: the 2010 February 19, March 4 and November 2 events. The top panels in Figure 1 present $^4$He, $^3$He, O, Fe intensities at 230--320\,keV\,nucleon$^{-1}$, the middle panels present He mass spectrograms at 0.4--10\,MeV\,nucleon$^{-1}$ and the bottom the inverted ion-speed spectrograms showing high energy ions arriving earlier than low energy ones. The corresponding solar electron events and the related type III radio bursts are shown separately in Figure \ref{fig:f2}. The top panels in Figure \ref{fig:f2} present electron intensities at different energy channels from Wind/3DP \citep{lin95} and ACE/EPAM \citep{gol98} instruments. The bottom panels are Wind/WAVES \citep{bou95} radio spectrograms. Note that Wind was also at L1 location. \added{The events shown in Figure \ref{fig:f1} have quite variable duration of $^3$He-rich SEP periods, ranging from $<$0.5\,day in the February 18 event to $>$2\,days in the November 2 event. Multiday periods ($\geq$2\,days) of $^3$He-rich SEPs (at 0.2--0.4\,MeV\,nucleon$^{-1}$) were observed in several events (2, 4, 6, 9, 12, 16, 21, 28, 29) of this study. Recurrent ion injections \citep{mas07} or a confinement of $^3$He in compression regions in solar wind \citep{koc08} have been considered as possible factors for such extended periods.}

\subsection{Solar sources} \label{subsec:ss}

\floattable
\begin{deluxetable}{cclcclccccccccl}
\rotate
\tabletypesize{\scriptsize}
\tablewidth{0pt} 
\tablecaption{L1 $^3$He-rich SEP events Jan 2007-- Dec 2010: Solar source properties \label{tab:tab2}}
\tablehead{
\colhead{Number} & \multicolumn{5}{c}{Flare} & \colhead{Type III} & \multicolumn{2}{c}{Electron event\tablenotemark{b}} & \multicolumn{2}{c}{CME\tablenotemark{c}} & \multicolumn{2}{c}{L1 footpoint} & \colhead{PFSS} & \colhead{Connection}\\
\cline{2-6}
\cline{8-9}
\cline{10-11}
\cline{12-13}
\colhead{} & \colhead{Class} & \colhead{Start} & \colhead{Location} & \colhead{NOAA AR} & \colhead{EUV type\tablenotemark{a}} & \colhead{start} & \colhead{3DP} & \colhead{EPAM} & \colhead{Speed} & \colhead{\ Width} & \colhead{Longitude} & \colhead{Time\tablenotemark{d}}& \colhead{polarity\tablenotemark{e}} & \colhead{to AR\tablenotemark{f}}
}
\startdata
1 & B2.6 & Jan 13 15:08 & S05W80 & 0933 & b & 15:12 & \nodata & 45--134 & \nodata & \nodata & W60 & \nodata & $(-)$ & yes\\
2 & B5.1 & Jan 24 00:28 & S05W58 & 0939 & b & 00:29 & 1.3--13 & 45--134 & \nodata & \nodata & W58 & \nodata & ($-$) & yes\\
3 & B3.9 & May 22 14:30 & N03W38 & 0956 & w (S) & 14:24 & 1.3--19 & 45--74 & 544 & 108 & W53 & 14:22--14:32 & (C) & \nodata\\
4 & A1.1 & Feb 4 13:35 & S11W24 & 0982 & b & \nodata & \nodata & \nodata & \nodata & \nodata & W41 & \nodata & ($-$) & yes\\
5 & \nodata & Jun 16 00:56 & S10W17 & 0998 & j & 00:55 & \nodata & \nodata & \nodata & \nodata & W37 & \nodata & ($-$) & yes\\
6 & C1.0 & Nov 4 03:17 & N37W46 & 1107 & w (NW) & 03:24 & 1.3--182 & 45--134 & 732 & 66 & W59 & 03:55--04:05 & ($-$) & yes\\
7 & \nodata & Apr 29 01:56 & N17W135 & \nodata & w (E) & 01:55 & 1.3--2.8 & \nodata & 239 & 44 & W75 & 02:45 & \nodata & \nodata\\
8 & B2.2 & May 1 08:54 & S09W85 & 1016 & w (S) & 08:57 & 1.3--182 & 45--134 & \nodata & \nodata & W76 & 09:05 & (C) & \nodata\\
9 & B3.6 & Jul 4 01:02 & S26E13 & 1024 & w (S) & 01:04 & \nodata & \nodata & \nodata & \nodata & W69 & 01:46 & (C) & \nodata\\
10 & \nodata &  Oct 6 08:10 & S31W102 & 1026 & b & 08:15 & \nodata & \nodata & \nodata & \nodata & W63 & \nodata & (+) & no (32$^{\circ}$)\\
11 & \nodata & Nov 3 03:30 & N12W130 & 1029 & w (SE) & 03:32 & \nodata & 45--235 & \nodata & \nodata & W59 & 04:15 & ($-$) & no (33$^{\circ}$)\\
12 & C7.2 & Dec 22 04:50 & S27W46 & 1036 & w (NE) & 04:54 & 1.3--4.2 & 45--235 & 318 & 47 & W64 & 05:05 & (+) & yes (2$^{\circ}$)\\
13 & \nodata & Jan 15 23:35 & N26W47 & 1040 & j & \nodata & \nodata & \nodata & \nodata & \nodata & W53 & \nodata & ($-$) & yes\\
14 & B3.2 &  Jan 26 17:01& N19W75 & 1042 & w (S) & 17:03 & 1.3--6.1 & 45--134 & \nodata & \nodata & W62 & 17:20 & (C) & \nodata\\
15 & \nodata & Jan 30 13:55 & N22W90 & \nodata & w & \nodata & \nodata & \nodata & 219 & 47 & W63 & 14:25 & \nodata & \nodata\\
16 & C7.7 & Feb 8 04:04 & N22W00 & 1045 & w (S) & 04:12 & 4.2--19 & 45--235 & \nodata & \nodata & W60 & 04:50 & ($-$) & no (17$^{\circ}$)\\
17 & M8.3 &  Feb 12 11:19& N25E12 & 1046 & w (S) & 11:25 & 13--19 & 45--235 & 509 & 360 & W68 & 12:00 & (+) & no (55$^{\circ}$)\\
18 & \nodata & Feb 18 18:58 & S19W58 & 1047 & w (SE), b & 19:03 & 1.3--19 & 45--134 & 223 & 14 & W51 & 19:03 & (C) & \nodata\\
19 & B6.0 & Mar 4 13:28 & S17W60 & 1052 & w (NE) & 13:30 & 2.0--4.2 & 45--235 & 374 & 44 & W51 & 13:35 & (C) & \nodata\\
20 & B1.4 & Mar 19 04:55 & N13W60 & 1054 & w (SW) & 04:49 & 1.3--19 & 45--134 & \nodata & \nodata & W56 & 04:55--04:58 & ($-$) & yes\\
21 & M2.0 & Jun 12 00:30 & N23W50 & 1081 & w (SW) & 00:54 & 6.1--182 & 45--235 & 486 & 119 & W64 & 01:05 & ($-$) & yes\\
22 & \nodata & Aug 31 20:50 & S22W145 & 1100 & w (NE) & 20:50 & 1.3--182 & 45--235 & 1304 & 360 & W72 & 21:25 & (C) & \nodata\\
23 & \nodata & Sep 2 10:48 & N24W62 & 1102 & j & 10:48 & 1.3--2.8 & \nodata & \nodata & \nodata & W54 & \nodata & ($-$) & yes\\
24 & B1.8 &  Sep 4 07:51& N23W85 & 1102 & b & 07:51 & 1.3--66 & 45--74 & \nodata & \nodata & W59 & \nodata & ($-$) & no (20$^{\circ}$)\\
25 & B5.7 & Sep 17 00:14 & S20W02 & 1106 & j & 00:16 & 40--182 & 45--74 & \nodata & \nodata & W49 & \nodata & (+) & yes\\
26 & C1.7 & Oct 17 08:52 & S18W32 & 1112 & w (NW) & 08:55 & 1.3--182 & 45--235 & 304 & 54 & W58 & 09:20 & (+) & yes\\
27 & C1.3 &  Oct 19 06:45& S18W58 & 1112 & b & 06:48 & 2.0--182 & 45--134 & 385 & 77 & W57 & \nodata & (+) & yes (8$^{\circ}$)\\
28 & B1.9 & Nov 2 07:26 & N19W90 & 1117 & w (SE) & 07:26 & 2.0--66 & 45 & 253 & 67 & W66 & 07:35 & ($-$) & yes (7$^{\circ}$)\\
29 & C1.1 & Nov 13 23:50 & S23W28 & 1123 & b & 23:50 & \nodata & \nodata & 442 & 63 & W49 & \nodata & ($-$) & yes\\
30 & B3.4 & Nov 17 08:07 & S23W72 & 1123 & j & 08:08 & 1.3--182 & 45--134 & 639 & 41 & W47 & \nodata & (+) & no (52$^{\circ}$)\\
31 & B2.4 & Nov 29 05:25 & N28E112 & 1131 & w (S) & 05:31 & \nodata & \nodata & 505 & 85 & W56 & 05:51 & \nodata & \nodata\\
32 & \nodata & Dec 10 05:55 & N35W99 & \nodata & w & \nodata & \nodata & \nodata & 299 & 72 & W69 & 06:20 & \nodata & \nodata\\
\enddata
\tablenotetext{a}{b: brightening, w: wave (direction of propagation), j: jet}
\tablenotetext{b}{energy ranges from Wind/3DP (1.3--182\,keV) and ACE/EPAM (45--235\,keV); background from a preceding event at low (events 1, 11) and high (events 7, 23) energies}
\tablenotetext{c}{speed (km\,s$^{-1}$), width ($^{\circ}$) from SOHO/LASCO catalog (\url{http://cdaw.gsfc.nasa.gov/CME\_list})}
\tablenotetext{d}{when the wave front intersects the L1 foot-point; for events 7, 9, 11, 17, 22, 31, 32 - when the wave front disappears}
\tablenotetext{e}{the source surface altitude at 2.5\,$R_{\sun}$; ($-$) negative, (+) positive polarity, (C) closed field}
\tablenotetext{f}{\replaced{if "no" then a distance (lat. or}{approximate} longitudinal \replaced{, whichever is farther)}{separation} between the L1 foot-point and open ecliptic field lines is indicated\deleted{when the PFSS and IMF polarities agree}}
\end{deluxetable}

The solar sources for the investigated $^3$He-rich SEP events were identified on the basis of their high association with type III radio bursts which are excited by low-energy electrons ($<$30\,keV) streaming outward through interplanetary space \citep[e.g.,][]{rea86, kru99}. The parent electrons are not always observed which could be an instrumental or magnetic connection issue. A similar identification approach has been used in previous studies \citep[e.g.,][]{nit06}. We first determined the approximate ion injection time at the Sun. For the events with a velocity dispersive onset (12 of 32) we extrapolated backward in time using the inverse velocity-time spectrograms \citep[e.g.,][]{rea85} and for the other events with unclear (9) or non-dispersive onset (11) we subtracted the interplanetary propagation time (without scattering) along the Parker spiral from the observed start time of the event. The uncertainty in the $<$1\,MeV\,nucleon$^{-1}$ ion injection time of $^3$He-rich SEP events has been estimated at $\pm$45 minutes using the velocity dispersion technique \citep{mas00}. The fitting of the heavy ion count-rate time profiles at several low-energy bins between 0.097 and 1.67\,MeV\,nucleon$^{-1}$ in ten good electron/$^3$He-rich SEP events provides the average uncertainty of $\pm$37 minutes \citep{wan16}. In the next step we inspected the Wind/WAVES dynamic radio spectra for \added{the events associated} type III radio emissions\deleted{around the inferred ion injection time}. \added{The type III bursts were observed in 29 events. In 19 cases (which include all dispersive events) the type III bursts were found in a five-hour window preceding the inferred ion injection time; in 24 cases the bursts occurred in ten-hour window.} The frequency range of the WAVES instrument covers emissions generated from about 2\,$R_{\sun}$ to 1\,AU. Then we localized the solar sources by examining high cadence (minimum 2.5-min resolution) full-disk STEREO EUV images from the SECCHI/EUVI instrument \citep{how08} for \replaced{a brightening/jet}{flaring} at the time of the type III burst. The EUVI observes the Sun in 2048$\times$2048 1.6 arc second pixels in a circular field of view which extends to 1.7\,$R_{\sun}$. During the very quiet period in 2007--2009 the source identification, even for the threshold events, was quite reliable as there was no simultaneous activity on the solar disk. In 2010, when the solar activity increased and the state of the corona became more complex, we benefited from the angular separation between the two STEREOs observing almost the full Sun (see Appendix for the concrete examples). STEREO-A is leading and STEREO-B trailing the Earth on their heliocentric orbits at $\sim$1\,AU. Two more radio receivers on the two STEREOs \citep{bou08} provided additional clues to the source location.  

\begin{figure*}
\figurenum{3}
\plotone{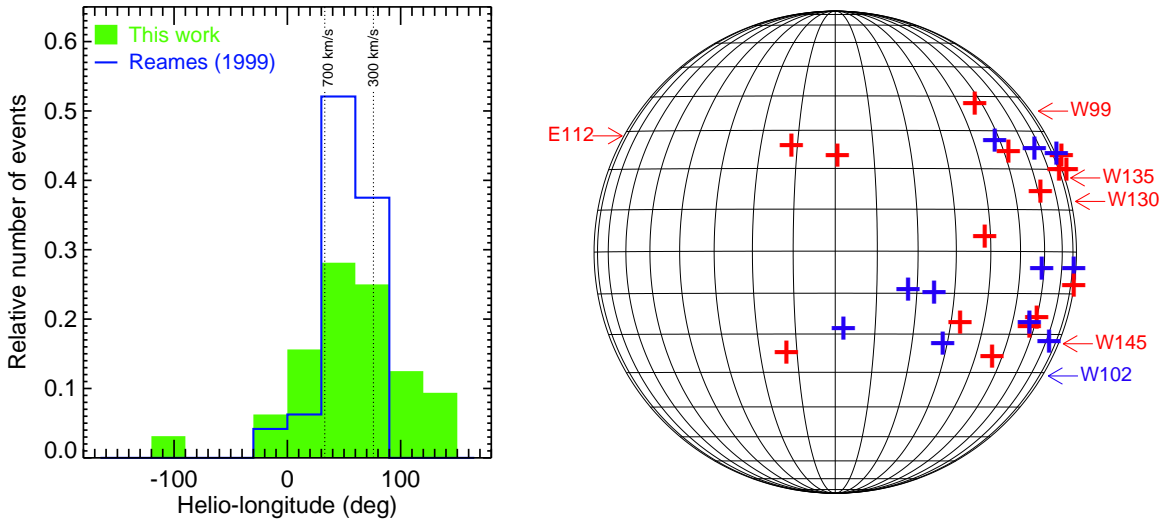}
\caption{Left: Histograms of flare longitudes associated with $^3$He-rich SEP events from this work (green shaded) and from \citet[][blue line]{rea99}. Two vertical dotted lines indicate L1 foot-point longitudes for two extreme solar wind speeds. Right: Pluses indicate the flare locations on the solar disk from the Earth view. Red pluses mark events associated with a coronal wave. The arrows point to the limb occulted flares. \label{fig:f3}}
\end{figure*}

Table \ref{tab:tab2} presents the solar source properties for the $^3$He-rich SEP events in this survey. Column 1 gives the event number. Column 2 indicates the GOES X-ray flare class and column 3 the flare start time from the Solar Event List (\url{ftp://ftp.swpc.noaa.gov/pub/warehouse}) but inputs in event 4 are from \citet{mas09}. If the X-ray flare is not reported then the time of the STEREO/EUVI flare is shown. Note that flares in events 2--6, 8, 12, 16, 17, 19, 21 are included in the STEREO/SECCHI EUVI catalog (\href{http://secchi.lmsal.com/EUVI/MOVIES\_FLARES\_CMES}{http://secchi.lmsal.com/EUVI/MOVIES\_FLARES} \href{http://secchi.lmsal.com/EUVI/MOVIES\_FLARES\_CMES}{\_CMES}) reporting flares in 2006 December--2010 October. Column 4 gives the flare location from the Earth view, column 5 the NOAA active region (AR) number, column 6 the EUV flare type (brightening, jet or wave). For events 3, 6, 12, 14 and 21 the EUV wave speed has been reported and ranges between $\sim$300\,km\,s$^{-1}$ and $\sim$500\,km\,s$^{-1}$\citep{war11,nit13,nit14,buc15}. Column 7 gives the type III radio burst start time as observed by Wind/WAVES, columns 8 and 9 the energy ranges of the solar electron event from the Wind/3DP (column 8) and ACE/EPAM (column 9) telescopes. The high-energy (40--182\,keV) electron observations from the Wind/3DP/SST are not available for events 1--3, 10--20. Columns 10 and 11 indicate the CME speed and width, respectively, obtained from the SOHO LASCO CME catalog. In events 9, 11, 16, 17 a CME is not reported but the coronagraph COR-1 on STEREO observed an ejection (in events 11 and 17 only faint outflow). The SOHO LASCO C2 and STEREO COR-1 cover 1.5--6\,$R_{\sun}$ and 1.4--4\,$R_{\sun}$, respectively. Column 12 gives the IMF foot-point longitude at a distance of 2.5\,$R_{\sun}$ from the Sun center, connecting L1 at the flare onset time. The L1 foot-point longitude was determined from the Parker spiral angle using the measured solar wind speed. Column 13 gives the time when the wave front crossed the foot-point location or disappeared. Column 14 indicates the magnetic field polarity of the open coronal field in the source AR. The polarity is not determined for four ARs (events 7, 15, 31, 32) which were not visible or present in the magnetograms. The AR in events 7, 15 was just emerging near the west limb and in event 31 it was behind the east limb. The last column indicates whether the L1 foot-point is connected to the source AR via coronal field. The coronal field is computed with a potential-field source-surface (PFSS) model \citep{sch69} using the Solar Soft PFSS package (\url{http://www.lmsal.com/~derosa/pfsspack}). The source surface which separates open and closed fields was set at 2.5\,$R_{\sun}$ from the Sun center (but see Section \ref{subsec:mc}). The appendix provides further features of the investigated events, particularly those with multiple ion injections and where the determination of a solar source was less straightforward. 

Figure \ref{fig:f3}, left panel, shows distributions of flare (X-ray or EUV) longitudes for the events in this survey and from an earlier study \citep{rea99} where source flares were determined from single spacecraft observations. The Gaussian fits to these distributions provide standard deviations of $\sim$42$^{\circ}$ and $\sim$16$^{\circ}$, respectively. As probably expected, with STEREO-A we are able to locate $^3$He-rich SEP sources behind the west solar limb. The right panel shows the flare distribution on the Sun's disk from the Earth view. Interestingly, almost all sources behind the limb (5 of 6) are associated with a coronal wave. The one with a longitude W145 was related to a high-energy solar proton event. 

\begin{figure*}
\figurenum{4}
\plotone{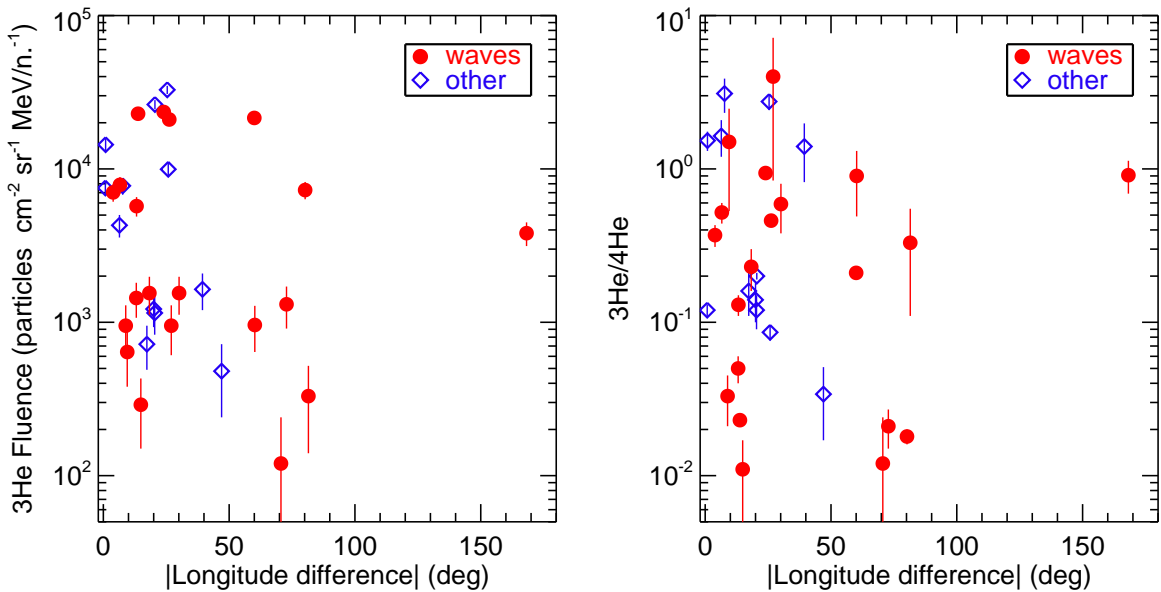}
\caption{$^3$He fluence (left) and $^3$He/$^4$He ratio (right) at 320--450\,keV\,nucleon$^{-1}$ versus a difference between the flare and L1 foot-point longitudes. Red circles mark the events with coronal waves and blue diamonds all remaining events in this work. The plots include also the event (number 11) where the fluence and ratio with large errors are not indicated in Table \ref{tab:tab1}. Recall the much lower $^3$He/$^4$He$\sim4\times10^{-4}$ in the solar wind \citep{glo98}. \label{fig:f4}}
\end{figure*}

Figure \ref{fig:f4} shows $^3$He fluence (left panel) and $^3$He/$^4$He ratio (right panel) versus longitudinal distance of L1 foot-point to the source flare. We can see that neither fluence nor ratio is obviously organised by this distance. These scatter plots show that source flares separated by $>$50$^{\circ}$ from a nominal connection  are all wave associated (two were with LSEP component). The median of $^3$He fluences in events associated with a coronal wave is slightly lower (a factor of $\sim$3) compared to the events with no waves. Right panel of Figure \ref{fig:f4} reveals that the $^3$He/$^4$He ratio is more scattered in the events with waves. Specifically, the events with lower ratios are those with a coronal wave but note that in three of the seven events with $^3$He/$^4$He $<$0.05 a high-energy solar protons were detected. The $^3$He/$^4$He (320--450\,keV\,nucleon$^{-1}$) median values are similar for both sets of events with ($^3$He/$^4$He$\sim$0.3) or without ($^3$He/$^4$He$\sim$0.2) coronal waves.   

\subsection{Coronal waves} \label{subsec:cw}

\begin{figure*}
\figurenum{5}
\plotone{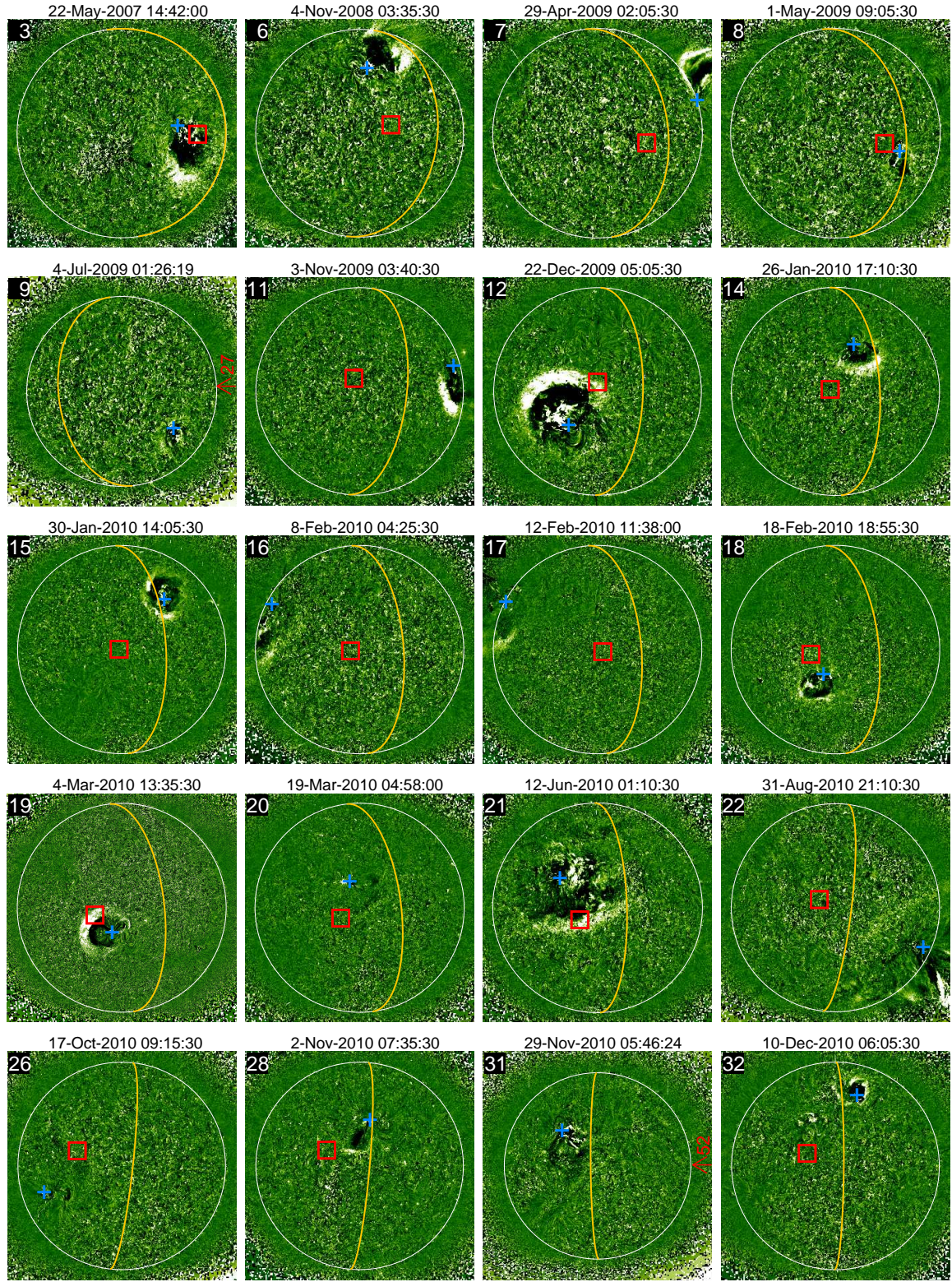}
\caption{STEREO 195\,{\AA} EUV running difference images of the solar disk. The 4-Jul-2009 and 29-Nov-2010 images are from STEREO-B; all others are from STEREO-A. Yellow arcs indicate the solar limb from the Earth view - the west limb in STEREO-A images and the east limb in STEREO-B. Red squares mark the L1 magnetic foot-point. Two red arrows (in the STEREO-B images) point to the foot-point indicating its longitude behind the limb. Blue pluses depict the source flare location. (Animations [\href{https://drive.google.com/open?id=0B4aWKR06L9wSTmlfcjJiSWtKWFU}{29-Apr-2009}, \href{https://drive.google.com/open?id=0B4aWKR06L9wSTlhJdmtDR19VdUE}{1-May-2009}, \href{https://drive.google.com/open?id=0B4aWKR06L9wSdVVUMVVDOUN0UzQ}{4-Jul-2009}, \href{https://drive.google.com/open?id=0B4aWKR06L9wSdVdXSTZXSnFkbUk}{19-Mar-2010}, \href{https://drive.google.com/open?id=0B4aWKR06L9wSNm5Ya1FRSXBnSlU}{31-Aug-2010}, \href{https://drive.google.com/open?id=0B4aWKR06L9wSdXc4ZGQ1a3JXa2M}{17-Oct-2010}, \href{https://drive.google.com/open?id=0B4aWKR06L9wSQjZOUUxWVkNlTzA}{29-Nov-2010}, \href{https://drive.google.com/open?id=0B4aWKR06L9wSVFpYMWtjVVlXakk}{10-Dec-2010}] are available.) 
\label{fig:f5}}
\end{figure*}

Figure \ref{fig:f5} shows the $^3$He-rich SEP event sources associated with coronal waves in the STEREO 195\,{\AA} EUV running differences images of the whole solar disk. The 195\,{\AA} channel observes emissions from a bright Fe~XII line formed at 1.6\,MK and a Fe~XXIV line formed at 20\,MK. The waves are visible as moving bright fronts of increased EUV emission followed by a dark area of decreased emission. The waves were poorly seen in the 171\,{\AA} EUV channel which however had a higher temporal cadence (2.5 minutes) during the first months of the STEREO mission. The images (mostly from STEREO-A) are shown at times, usually a few minutes after the X-ray flare, when the coronal waves were well visible. The waves associated with $^3$He-rich SEPs in events 14 and 28 have been included in recent studies \citep{nit15,buc15}. The most spatially extended wave fronts were seen in the $^3$He-rich events with high-energy solar proton population (12, 21, 22). The bright fronts were observed in the 2008 November 4, 2009 November 3, 2010 January 27 and 2010 March 4 $^3$He-rich events. It has been noted that bright and sharp fronts may indicate shocks \citep[e.g.,][]{bie02} \replaced{and that the shocks}{which} may be formed in the early stages after the wave expulsion \citep[e.g.,][]{war11}. The type II radio bursts, a coronal shock signature, are reported for three events (3, 12, 21). Two events (2010 March 19, 2010 October 17) show no clear waves but the motions of a dimming front can be seen in the animations. It might be that with 5-min resolution images we miss brighter wave fronts in these events. The wave fronts in the events of 2009 May 1 and 2009 July 5 are smaller in size and less bright. We observe that several wave-associated events (e.g., 7, 14, 19, 20) started with a jet. The animations, shown in Figure \ref{fig:f5} for faint wave events, contain only a few frames to demonstrate the wave front motions. In the 2009 April 29 and May 1 events we have also limited EUV data coverage.\deleted{Note that in half of the events with a coronal wave the source region was limb occulted or located near the limb as viewed from the Earth. In the remaining 10 events with the source visible or partly visible from the Earth, three are associated with a high energy proton component and three are associated with faint wave fronts.}

The analysis of sequences of EUV difference images showed that the wave fronts in seven $^3$He-rich SEP events propagated toward the L1 foot-point position but faded before crossing that point. \replaced{Specifically in}{In} the following four events, 2009 April 29 \added{(number 7)}, 2009 November 3 \added{(11)}, 2010 February 12 \added{(17)} and 2010 September 1 \added{(22)}, the waves \replaced{approach quite close to the L1 foot-point, disappearing}{disappear} $\sim$35--50 minutes after the type III radio burst onsets. It may be only an instrument resolution issue that they are not seen crossing the foot-point. The remaining three events, 2009 July 5 \added{(9)}, 2010 November 29 \added{(31)} and 2010 December 10 \added{(32)}, do not show significant lateral expansion of the associated waves.\deleted{Here it might be possible that the waves crossed (or approached) the coronal field lines which were connected to the L1 foot-point. But this is unexpected for the November 29 event where the angular distance to the foot-point is extraordinary large. Figure \ref{fig:f6} in the next section demonstrates that the coronal field lines can easily spread out 50$^{\circ}$--60$^{\circ}$ in longitude \citep[see also][]{kle08,wie13}. The connection may occur also at higher altitudes in the corona for the waves with dominant radial expansion \citep{lar14}. But not a particularly wide CME (85$^{\circ}$) was associated with the November 29 event.}In all other 13 $^3$He-rich events the waves passed the L1 foot-point (see Table \ref{tab:tab2}, Column 13). For a quite remote source AR in the 2008 November 4 \added{(6)} event the west flank of \replaced{the wave}{a dim wave front} passed by the L1 foot-point at $\sim$04\,UT (35 min after type III burst)\replaced{, in}{. In} the 2010 January 31 \added{(15)} event \added{the wave crossed the foot-point} at $\sim$14:25\,UT and in the 2010 February 8 \added{(16)} event at $\sim$04:50\,UT (38 min after type III emission). In the remaining 10 events delays were shorter, mostly $\sim$5--15 minutes which is within the ion injection time uncertainties. Note that all these events with larger wave front delays are too weak for performing any reliable timing analysis. Several previous works reported delayed ion injection by about 1\,hr relative to the solar electrons in some $^3$He-rich SEP events \citep[e.g.,][]{rea85,wan16}. This may be consistent with the above mentioned later wave front arrivals at the L1 foot-point. 

\subsection{Magnetic connection} \label{subsec:mc}

\begin{figure}
\figurenum{6}
\plotone{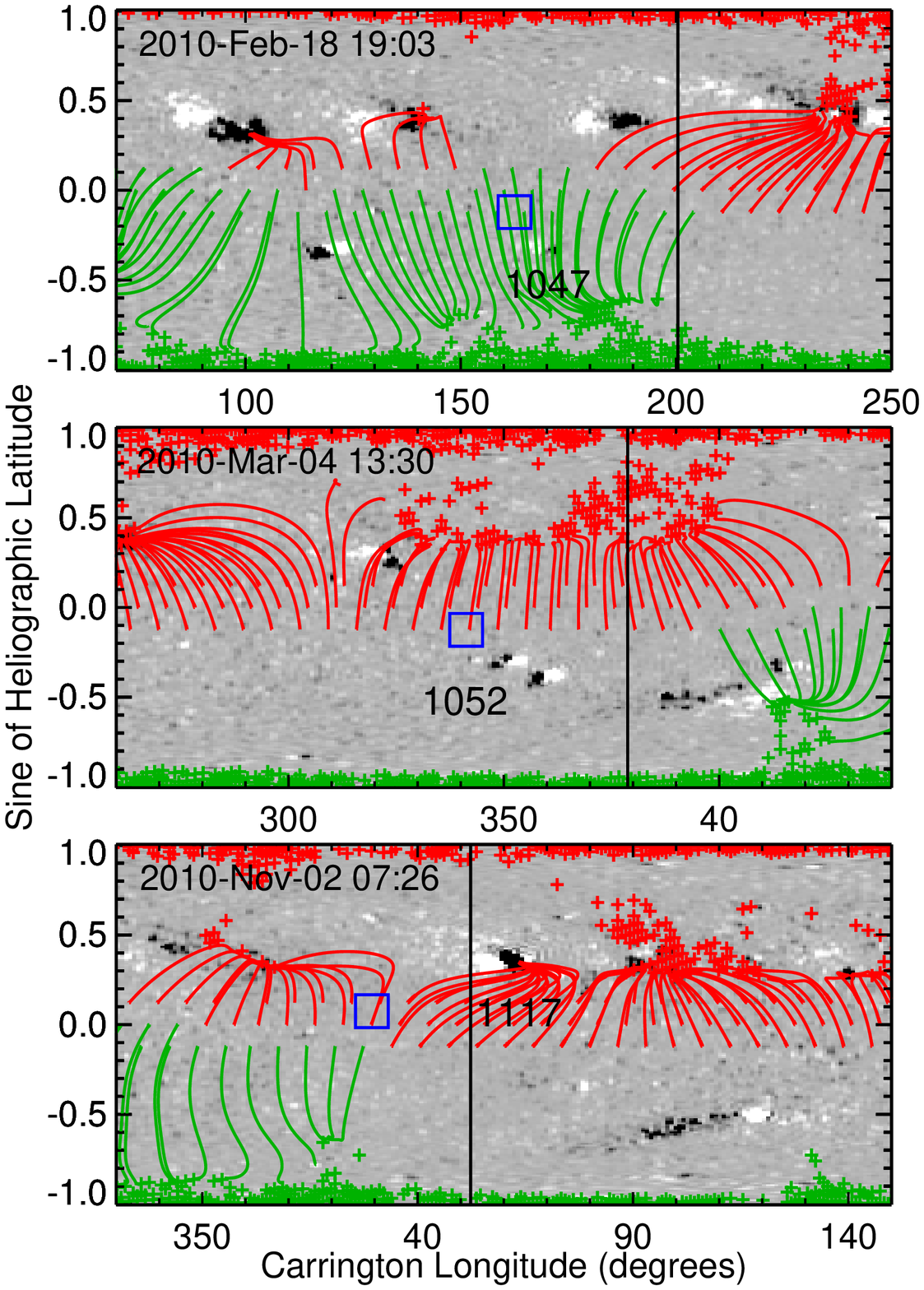}
\caption{Photospheric magnetic field (gray scale) and PFSS model coronal field (red - negative and green - positive polarity) for the 2010 February 19, March 4 and November 2 events around the type III radio burst onset times. Shown are field lines which intersect source surface at latitudes 0$^{\circ}$ and $\pm$7$^{\circ}$. Squares mark L1 magnetic foot-point at the source surface. Black vertical lines mark the west solar limb from the Earth view. The event source ARs are indicated by NOAA number. \label{fig:f6}}
\end{figure}

Figure \ref{fig:f6} shows photospheric magnetic field maps with the PFSS model coronal field (with source surface at 2.5\,$R_{\sun}$) for the \replaced{three}{wave-associated} events (18, 19, 28) from Figure \ref{fig:f1}. The source ARs, marked by NOAA number, show no open field in the February 19 and March 4 events. As seen in the magnetograms these two \replaced{sources}{ARs} are quite faint, having small size and no particularly strong magnetic fields. \replaced{The source AR in the February 19 event}{One region (AR1047)} is not well seen behind the set of open field lines emanating from the south-polar region. A further common feature of these two events is a close proximity of the L1 foot-point to the source AR \added{and thus easily reachable by the expanding wave}. \added{The L1 foot-point in the November 2 event could be connected to AR1117. The foot-point is separated by $\sim$7$^{\circ}$ in longitude from the open ecliptic field lines emanating from the source AR. This is within a 10$^{\circ}$ uncertainty of the foot-point location based on a simple Parker spiral \citep{nol73}.} The polarity of the \added{extrapolated} open field\deleted{in the source AR}for the November 2 event is inconsistent with the in-situ polarity. Note that in this case the magnetic field in the AR located close to the solar limb is not well observed.

Tracing the magnetic field through interplanetary space and the corona \citep[e.g.,][]{neu98}, from L1 to its origin at the Sun, was performed for almost all events in this study. For events 7, 15, 31, 32 where the associated ARs are not visible in the magnetograms the coronal field could not be extrapolated. \added{The magnetic connection would be possible for events 15 and 32, where the L1 foot-point is separated by $\sim$30$^{\circ}$ in longitude from the source AR. It is unexpected for event 31 where the angular distance to the foot-point is extraordinary large (170$^{\circ}$). Figure \ref{fig:f6} demonstrates that the coronal field lines can spread out 50$^{\circ}$--60$^{\circ}$ in longitude \citep[see also][]{kle08,wie13}. These four events were associated with a coronal wave which crossed the foot-point in event 15 and maybe in event 7. The connection may occur also at higher altitudes in the corona for the waves with dominant radial expansion \citep{lar14}. But not a particularly wide CME (85$^{\circ}$) was seen in event 31.} 

We find the source ARs in seven events (3, 8, 9, 14, 18, 19, 22) showing no-open field configuration according to the PFSS model. Interestingly, all these events were associated with an EUV wave \added{which clearly passed by the foot-point in five cases}. By decreasing the source surface altitude to 1.9\,$R_{\sun}$ the AR in the two events (14, 18) already showed open field (consistent with the in-situ polarity) connected to the L1 foot-point in event 18\deleted{and within 12$^{\circ}$ of latitude (and $\sim$5$^{\circ}$ of longitude) from L1 foot-point in event 14}. Note that in some events the associated AR was in unobserved (event 22) or not well observed (8, 14) areas behind or near the west limb, where the magnetograms are based on flux transport models. 

The source ARs in another seven events (2, 10, 17, 20, 27, 28, 30) show open to ecliptic coronal field with polarity inconsistent with the in-situ polarity. The events 17, 20, 28 were associated with an EUV wave that could \replaced{cause particle transport}{distribute particles} on the field with opposite polarity. \added{Indeed, the L1 foot-point was passed by the wave in events 20 and 28.} Note that the source ARs in events 10 and 28 were in unobserved areas, near the west solar limb (at W102 and W90, respectively). The L1 (or ACE) foot-point in the \replaced{event 2}{events 2, 20, 27, 28} is connected to the source AR\added{, though} via coronal field with the polarity different from the in-situ. The polarity would be consistent if source surface is lowered to 1.9\,$R_{\sun}$ for events 27 and 30. Then the L1 foot-point \replaced{is connected to the source AR}{shows connection} in event 27 and it is within 20$^{\circ}$ of longitude \added{from the open field lines} in event 30.

The ACE foot-point at the source surface was directly connected via coronal field (consistent with IMF) to the associated AR in \replaced{eight}{ten} events (1, 4, 5, 6, \added{12,} 13, 21, 23, \added{26,} 29)\deleted{and in two other cases (12, 26) the ACE foot-point was within 5$^{\circ}$ of latitude (and $\sim$0$^{\circ}$ of longitude) from the coronal field lines connecting the source AR}. \added{Recall the events 6, 12, 21 and 26 were associated with an EUV wave.} In three other events (11, 16, 24) the ACE foot-point was within 33$^{\circ}$, 17$^{\circ}$ and 20$^{\circ}$ of longitude from the coronal field connecting the source AR. Two of them (11, 16) were associated with an EUV wave \added{which crossed the foot-point in event 16 and maybe also in event 11}. \added{Notice that the source AR in event 11 was behind the limb (W130) and the coronal field might not be correctly extrapolated.} The source AR in event 25 shows open to ecliptic field lines with both polarities, where the ACE foot-point is\deleted{directly} connected\deleted{to the source} via the positive polarity field \added{(inconsistent with IMF)} and it is 40$^{\circ}$ of longitude from the negative polarity field lines. \deleted{Note there is a $\pm$10$^{\circ}$ uncertainty in the foot-point location based on a simple Parker spiral \citep{nol73}.}

\added{\subsection{Multi-spacecraft $^3$He-rich SEP events} }\label{subsec:mse}

\added{Table \ref{tab:tab3} shows $^3$He-rich SEP events of this survey observed on angularly separated STEREO-A (STA) or -B (STB) spacecraft \citep{wie10,wie13}. Interestingly, all these events are associated with EUV waves. Column 1 gives the ACE event number. Columns 2, 3 and 4 indicate for STB an observation of the event, the foot-point longitude at the source surface and the foot-point connection to the source AR, respectively. Columns 5-7 indicate the same for STA. Columns 8 and 9 give the spacecraft separation angles from the Earth. The STEREO $^3$He-rich events were identified using the LET \citep{mew08} instrument in the range 2.3--3.3\,MeV\,nucleon$^{-1}$. The authors excluded chance coincidences between detections of $^3$He-rich SEPs from widely spaced ARs.}

\added{Event 6 was observed on all three spacecraft. STB and ACE were connected to the event associated AR. STA was not but its coronal foot-point was crossed by the EUV wave. Event 7 was seen on ACE and STA. Though no field extrapolations are available, the STA foot-point (separated by 17$^{\circ}$ in longitude) might have been connected to the source AR. The ACE foot-point, too far (60$^{\circ}$) for a direct connection, was maybe reached by the EUV wave (see Section \ref{subsec:cw}). Event 9 was observed on STB and ACE. The spacecraft foot-points, quite separated from the event related AR with a closed field configuration, were not seen to be crossed by the wave. Event 11 was observed on ACE and STA where only STA was connected. The ACE foot-point was probably reached by the wave. Events 16 and 17 were seen on all three spacecraft with only STB connected to the source AR. In event 16 the EUV wave passed by the ACE foot-point but no the STA; in event 17 none of these two foot-points were passed by the wave. Event 21 was measured on STB and ACE but only ACE was connected to the associated AR. The STB foot-point was passed by the wave. The source AR in event 22 (seen on ACE and STA) shows closed field. The EUV wave passed by the STA and likely also the ACE foot-point. Event 28 was observed on ACE and STA. While the ACE foot-point was connected and reached by the wave, the STA was not. }

\added{
\begin{deluxetable*}{ccllcllcc}
\tablecaption{Angle-separated $^3$He-rich SEP events \label{tab:tab3}}
\tablehead{
\colhead{Number} & \multicolumn{3}{c}{STEREO-B} & \multicolumn{3}{c}{STEREO-A} & \multicolumn{2}{c}{Separation}\\
\cline{2-4}
\cline{5-7}
\cline{8-9}
\colhead{} & \colhead{Event} & \colhead{Footpoint\tablenotemark{a}} & \colhead{Connection\tablenotemark{b}} & \colhead{Event} & \colhead{Footpoint\tablenotemark{a}} & \colhead{Connection\tablenotemark{b}} & \colhead{STB-Earth} & \colhead{Earth-STA}
}
\startdata
6 & yes & W31 & yes & yes\tablenotemark{c} & W80 (wf) & no (20$^{\circ}$) & 41$^{\circ}$ & 41$^{\circ}$ \\
7 & no & W15 & \nodata & yes & W118 (wf) & \nodata & 47$^{\circ}$ & 48$^{\circ}$ \\
9 & yes & W32 & \nodata & no & W103 & \nodata & 48$^{\circ}$ & 54$^{\circ}$ \\
11 & no & W15 & no (76$^{\circ}$) & yes & W119 (w) & yes & 61$^{\circ}$ & 63$^{\circ}$ \\
16 & yes\tablenotemark{d} & E13 (w) & yes (2$^{\circ}$) & yes & W112 & no (70$^{\circ}$) & 71$^{\circ}$ & 65$^{\circ}$ \\
17 & yes & W10 (w) & yes & yes & W111 & no (95$^{\circ}$) & 71$^{\circ}$ & 65$^{\circ}$ \\
21 & yes & E23 (wf) & no (30$^{\circ}$) & no & W121 & no (36$^{\circ}$) & 70$^{\circ}$ & 74$^{\circ}$ \\
22 & no & W08 & \nodata & yes & W126 (w) & \nodata & 74$^{\circ}$ & 81$^{\circ}$ \\
28 & no & E01 & no (70$^{\circ}$) & yes & W130 & no (25$^{\circ}$) & 82$^{\circ}$ & 84$^{\circ}$ \\
\enddata
\tablenotetext{a}{an intersection with the wave (w) or the wave flank (wf) is indicated}
\tablenotetext{b}{the longitudinal separation between the foot-point and open ecliptic field lines is indicated}
\tablenotetext{c}{close to detection threshold}
\tablenotetext{d}{associated with an earlier injection than the event on ACE}
\end{deluxetable*}
}

\section{Discussion and conclusion}\label{sec:con}

We collected 32 $^3$He-rich SEP events observed by ACE at L1 during the solar minimum period 2007--2010. This extended period of low solar activity provided favorable conditions for the identification of $^3$He-rich solar sources characterized by faint flare signatures. We include in this study also weak events with $^3$He intensity close to the detection threshold. We examined the solar sources with STEREO EUV imaging observations. At the beginning of 2007 the STEREO-A was near the Earth while at the end of the investigated period the angular separation extended to $\sim$90$^{\circ}$ from the Earth-Sun line allowing for the first time a direct view of the near west solar limb SEP sources. 

Surprisingly, we find that more than half of the events of this survey (20 of 32) are clearly associated with large-scale coronal waves. The finding remains unchanged (16 of 28) after excluding the events with a high-energy solar proton component. Other events, not counted in this number, like the 2008 June 16 and 2010 November 14 events, might also have secondary ion injections related to the wave expulsion (see Appendix). The EUV wave fronts passed by or showed close approach the L1 magnetic foot-point in several ($\sim$13--17) investigated events. However, this occurred either too fast after the type III radio burst onset or the events were too weak to distinguish between the flare and the potential wave-front ion injections. At least 10 events of this study have been associated with EUV jets including the wave events starting as a jet. Some events classified as having a brightening might have shown a narrow ejection from a different observing angle. Several $^3$He-rich events in the investigated period were reported in previous surveys (see references in Table \ref{tab:tab1}). \citet{mas09} focused on solar activity minimum conditions associated with acceleration of $^3$He-rich SEPs during the period March 2007--December 2008, \citet{wie13} on $^3$He-rich SEPs with wide longitudinal spread observed on multiple spacecraft between January 2007 and January 2011, and \citet{nit15} on solar source characteristics derived from high resolution EUV images on SDO in the period May 2010--May 2014. 

The present study shows a quite broad source flare longitude distribution in the $^3$He-rich SEP events where the most remote locations were associated with EUV waves. It has been noted that the connection to a distant source can not be sufficiently explained by a divergence of the coronal field lines below the source surface \citep{wie13}. The magnetic connection controlled by large scale EUV waves in the corona has been discussed in several LSEP events \citep[e.g.,][]{tor99,kru99,mal09,rou12,nit12,par13,ric14} and recently suggested as a possible mechanism also for $^3$He-rich SEPs \citep{nit15}.\deleted{We found that all previously reported events with a wide longitudinal spread of $^3$He-rich SEPs are associated with large-scale coronal waves. This provides new insights on energetic ion transport from their sources on the Sun to 1 AU.}\added{We examined magnetic connection combining the PFSS extrapolations of coronal field with Parker spiral model of IMF.} The PFSS model has been considered to be adequate for the extrapolation of global magnetic fields in the corona. For our 12 $^3$He-rich SEP events with jet/brightening the\deleted{PFSS model (with the source surface at 1.9--2.5\,$R_{\sun}$) predicts} magnetic connection \replaced{via coronal field consistent with the in-situ polarity in 10 cases (events 1, 4, 5, 13, 23, 24, 25, 27, 29, 30) and fails only in 2 cases (events 2, 10).}{to the source AR is found in 9 cases (events 1, 2, 4, 5, 13, 23, 25, 27, 29) while} for 16\deleted{$^3$He-rich SEP} events with a coronal wave \replaced{, the magnetic connection is correctly predicted only for 8 cases (6, 11, 12, 14, 16, 18, 21, 26); in the remaining 8 events the PFSS model shows closed field  (3, 8, 9, 19, 22) or open field that is inconsistent with the IMF polarity (17, 20, 28). The remaining 4 events with EUV waves (7, 15, 31, 32) have the AR not visible in the magnetograms}{only in 6 cases (6, 12, 20, 21, 26, 28)}. In addition none of the events with jets show closed model field but all cases with a closed configuration are found for the events with EUV waves. \added{Thus the connection to the flaring region on the Sun was generally not required for the events with a coronal EUV wave.} \replaced{Thus the model field extrapolation, with its known limitations, also}{This} suggests that in some $^3$He-rich SEP events the ion injection may be instigated by EUV wave expulsions.

We found that \replaced{all previously reported}{9} events \added{in this study} with a wide longitudinal spread of $^3$He-rich SEPs are associated with large-scale coronal waves. \added{In six angle-separated events (6, 7, 11, 16, 21, 22) the spacecraft coronal foot-points without magnetic connection to the source AR were traversed by EUV waves, probably ensuring an injection of SEPs onto IMF lines. In the remaining three events (9, 17, 28), where EUV waves were not seen crossing the foot-points, the wide angular distributions could be due to other mechanizms. These have been thoroughly discussed elsewhere \citep[e.g.,][]{wie13,ric14} and may include particle cross-field diffusion (in corona or interplanetary space) and a distortion of magnetic fields by CMEs.}


In summary, the multi-spacecraft EUV imaging and radio observations on angularly separated STEREOs provided the most reliable identification of $^3$He-rich SEP sources as we have ever had. This study reveals quite common ($\sim$60\%) association of $^3$He-rich SEPs with large-scale coronal EUV waves. While in LSEP events the presence of significant CMEs expanding in all directions can mimic the role of EUV waves, the relation with energetic ions can be better seen in $^3$He-rich SEP events which are often found without a CME or accompanied by weak coronal outflows. Note that half of the EUV wave events in this survey are accompanied by slow CMEs ($\lesssim$300\,km\,s$^{-1}$) or by no CME at all. 

\acknowledgments
The work of R. Bu\v{c}\'ik is supported by the Deutsche Forschungsgemeinschaft (DFG) under grant BU 3115/2-1. We thank R.~J. MacDowall for providing the Wind/WAVES data and acknowledge the use of the Wind/3DP data. Work at JPL was supported by NASA. Work at APL (ACE/ULEIS and STEREO/SIT) was supported by NASA grant NNX13AR20G/115828 and NASA subcontract SA4889-26309 from the University of California Berkeley.

\appendix

Details of the key event observations are summarized below.\\

\paragraph{2007 January 24 event} 

This multiday (3 days) event was associated with a B5.1 flare (January 24 00:28\,UT) from AR0939 (see Table \ref{tab:tab2}). The same region produced $\sim$5 hrs later a B6.8 flare (05:13\,UT), a type III radio burst (05:15\,UT), and an electron event but no obvious new $^3$He increase. This later flare was associated with a CME (67$^{\circ}$ wide; 295\,km\,s$^{-1}$). The STEREO SECCHI EUVI catalog reports a jet-like ejection. The candidate B5.1 flare was shortly preceded by a B1.2 flare (January 23 23:58\,UT), a type III radio burst and a jet (in EUVI catalog) from the same AR. 

\paragraph{2007 May 23 event} 

The same AR (N03W53) launched later another EUV wave \citep[in EUVI catalog;][]{nit14} in association with a B5.3 flare (May 23 07:15\,UT), type III (07:17\,UT) and type II (07:22\,UT) radio bursts, a CME (679\,km\,s$^{-1}$, 90$^{\circ}$ wide) and a solar electron event. The high-energy solar proton event has been also reported \citep{ric14}.

\paragraph{2008 February 4 event} 

This weak event, reported in an earlier survey \citep{mas09}, has not been associated with a particular type III radio burst. The type III bursts that occurred closest in time to the SEP event were produced by the source AR ($\sim$W02) on February 3 (00:35, 03:23\,UT), less than one day before the start of the event. These were accompanied by jets but no electron events or X-ray flares. Though one day is certainly a long time for SEPs to take to arrive at L1, the events near the detection threshold may only show peak intensities and not the event onset or its early rising phase.  

\paragraph{2008 June 16 event} 

There are two possible candidate flares for this weak event: 1) a brightening in the source AR at 05:25\,UT (see EUVI catalog) with a faint wave-like motion accompanied by a weak type III burst (barely visible and only below 0.5\,MHz); the COR1 CME catalog (\url{http://cor1.gsfc.nasa.gov/catalog}) reports narrow ejection seen by STEREO-B; 2) a quite intense type III burst at 00:55\,UT, 20 hrs before the event appearance, with a jet-like ejection (Table \ref{tab:tab2}). Also in this weak event the observed start of $^3$He period may already correspond to the peak intensities. 

\paragraph{2008 November 4 event} 

The source AR (N35W75) produced another strong type III burst and an A9.3 flare (included in EUVI catalog) on November 6 at 11:19\,UT with a narrow (29$^{\circ}$) CME (487\,km\,s$^{-1}$) but no electron event was observed. Whether this flare provided $^3$He-rich SEPs to the end of this multiday period is not clear.

\paragraph{2009 April 29 event} 

The only active region (AR1016) on the solar disk seen from the two STEREOs (angularly separated by 95$^{\circ}$) appears to not be the event source. The type III radio burst preceding the event was stronger (over the whole frequency range) on STEREO-A than on Wind suggesting a more westward origin. In addition, at the time of the type III radio burst the AR1016 (S08W53) showed only a weak brightening. Simultaneously STEREO-A/EUVI observed a huge eruption behind the west solar limb that was most likely related to this event (see Table \ref{tab:tab2}). 

\paragraph{2009 July 5 event}

The event was observed on STEREO-B and at the threshold level on ACE \citep{wie13}. The solar electron event was only observed on STB (3.6--41\,keV) with the STE instrument \citep{lin08}. An association with the solar source is straightforward as there was only one AR on the solar disk as seen from the two separated STEREOs.

\paragraph{2010 January 16 event}

Though there was only one AR (AR1040) on the solar disk as observed from the two widely separated (134$^{\circ}$) STEREOs the source identification is not obvious. Since the event started at the beginning of January 16, the ion injection would be expected around the end of previous day. However, only one (and too late for the event onset) type III radio burst (January 16 05:41\,UT) was observed within the two day interval January 15 00:00\,UT--January 17 00:00\,UT. This originated in AR1040 and was accompanied by a jet-like emission. On January 15 there was little activity seen on the solar disk but a jet in AR1040 at 23:35\,UT would be a candidate source for this event. Though the source flare for the event can not be unambiguously determined there were nothing like EUV waves seen on the solar disk. 

\paragraph{2010 January 31 event}

The start time of $^3$He period of this weak event implies an ion injection before $\sim$16\,UT on January 30. On that day the Wind/WAVES data show three quite weak type III radio bursts: the burst at 10:47\,UT from the region behind the west limb in the STEREO-A (separated 65$^{\circ}$ from L1) view, and the bursts at 14:35\,UT (N30E125) and 17:55\,UT (N30E120) from the region behind the east limb in the Earth view. The most noticeable activity includes the expulsion of two coronal waves on January 30 13:55\,UT and 17:18\,UT (also a B1.1 flare) from a small AR located at the west limb (in the Earth view). The coronal waves propagated symmetrically to all sides and one at 13:55\,UT appears to be the most plausible candidate for this event. 

\paragraph{2010 February 12 event}

Note that another AR (AR1045, N20W70), associated with a previous (February 8) event, produced quite intense type III radio burst (February 12 17:05\,UT), a jet and a narrow EUV wave and might also have contributed later to the February 12 event. 

\paragraph{2010 February 19 event}

The event is associated with an EUV brightening, a type III radio burst (19:03\,UT) and an electron event \added{(see Figure \ref{fig:f2})}. However the same AR released $\sim$25 min earlier an EUV wave which was still visible in the vicinity of the AR at time of the type III burst onset. 

\paragraph{2010 June 12 event}

\citet{koz11} reported high initial speed (1287\,km\,s$^{-1}$) of the associated EUV wave and an open field geometry in the source AR. The authors suggested that this wave may be the EUV signature of a coronal shock. The same region (N23W52) produced 8 hrs later another flare (C6.1, 09:02\,UT), type III (09:12\,UT) and type II (09:16\,UT) emissions, an EUV wave \citep{nit13} and a CME (382\,km\,s$^{-1}$, 76$^{\circ}$ wide). The associated electron event is probably masked by the ongoing electron event from the previous flare. The $^3$He-rich SEP contribution from this flare to this multi-day ($\sim$3 days) event can not be ruled out. The same concerns a C1.5 flare from the same AR (N25W70) on June 14 00:41\,UT, accompanied by a type III radio burst (00:47\,UT), an electron event, a small-scale EUV wave (included in the EUVI catalog) and a CME (343\,km\,s$^{-1}$, 62$^{\circ}$ wide). 

\paragraph{2010 September 1 event}

The event was observed also on STEREO-A \citep{wie13}, separated 81$^{\circ}$ on the west from L1, together with a dispersive solar electron event observed with the STE and SEPT \citep{mul08} instruments. The L1 electron event was non-dispersive and delayed (by 25 min at 100\,keV) relative to the event on STA. The SECCHI/EUVI flare was very bright. 

\paragraph{2010 September 4 event}

The source AR (N23W90) produced $\sim$7 hrs later another type III burst (14:33\,UT), a B2.5 flare (14:31\,UT), jet-like emission and a dispersive solar electron event that also might have contributed to the observed $^3$He-rich SEPs. 

\paragraph{2010 October 19 event}

We note that $\sim$9 hrs later another flare (B3.2; 15:57\,UT) in the source AR, accompanied by an EUV brightening and a type III burst (but no solar electron event), could contribute to this event especially with low energy ($<$1\,MeV\,nucleon$^{-1}$) SEPs which started near the end of October 19 and peaked in the middle of next day. 

\paragraph{2010 November 2 event}

Note that the associated B1.9 flare at 07:26\,UT was closely followed by a B3.3 flare at 10:00\,UT, as well as an electron event \added{(see Figure \ref{fig:f2})} and a jet from the same AR. 

\paragraph{2010 November 14 event}

This is a multiday event where the $^3$He-rich SEP period lasted for 3 days. \citet{che15} pointed out that a B7.6 flare at 14:36\,UT on November 15 in the same AR, about one day after the beginning of this $^3$He-rich period, could contribute SEPs. The flare produced an electron event but no obvious $^3$He-rich SEP increase. It was associated with a fast eruption seen from STA and STB (in COR1 CME catalog) and the EUV wave \citep[][see catalog at \url{http://aia.lmsal.com/AIA\_Waves/}]{nit13}.

\paragraph{2010 November 17 event}

Note that during the event ($\sim$10 hrs after the associated flare) the source AR (S23W78) produced another EUV jet, a type III (18:13\,UT), and electron event (2--66\,keV).

\paragraph{2010 November 29 event}

Notice that the location of a $^3$He-rich SEP source region so far to the east (E112) appears to be quite peculiar. However, there are examples of $^3$He-rich SEP sources at similar eastern longitudes. \citet{wan12} have reported four electron- and $^3$He-rich SEP source ARs between E80--E60. More extreme locations might have been omitted in the past with imagers observing only Sun's front side. The candidate region for the November 29 event produced several strong type III radio bursts (01:19, 03:50,  05:31, 11:25, 19:05\,UT) as observed by STEREO-B. On STEREO-A all these were very faint, brief and not extending to the high frequencies. Two bursts at 05:31\,UT and 19:05\,UT were accompanied by a coronal wave. The start time of the $^3$He period implies ion injection before $\sim$8\,UT. STEREO-B, separated by $\sim$90$^{\circ}$ from L1 on the east, had better magnetic connection to this source. The higher energy $^3$He was not observed by LET (at 2.3--8.0\,MeV\,nucleon$^{-1}$) on STEREO-B, which is consistent with lack of high energy $^3$He at SIS. There was only a small He event below 1\,MeV\,nucleon$^{-1}$ detected by SIT \citep{mas08} on STEREO-B but $^3$He could not be resolved due to large instrumental background.  

\listofchanges
\end{document}